\documentstyle[12pt,epsfig]{article}

\title{%
\hfill
 \parbox{5cm}{%
 }\\
\vspace{5ex}
  T-violation search with very long baseline neutrino oscillation 
  experiments
\vspace{5ex}
}

\author{
 Masafumi Koike%
\thanks{e-mail address: {\tt koike@icrr.u-tokyo.ac.jp}}\\%
{\footnotesize \it%
 Institute for Cosmic Ray Research, University of Tokyo,
 Midori-cho, Tanashi, Tokyo 188-8502, Japan%
}
\\
and
\\
Joe Sato%
\thanks{e-mail address: {\tt joe@rc.kyushu-u.ac.jp}}\\%
 {\footnotesize \it%
   Research Center for Higher Education,
   Kyushu University,
   Ropponmatsu, Chuo-ku, Fukuoka
   810-8560, Japan%
 }%
}

\date{}

\begin{document}

\maketitle
\abstract{%
    We consider possibilities of observing T-violation effects in
    neutrino oscillation experiments with very long baseline ($\sim$
    10000 km) using low energy neutrino ($\sim$ several hundreds MeV). 
    We show that the matter effect effectively changes only the
    first-second generation mixing angle,
    respecting solar neutrino deficits and atmospheric neutrino
    anomalies.  The effective mixing of the first-second generation in
    the Earth grows up to maximum by resonance.  This effect enables
    one to search T violation in case the first-second mixing angle is
    small.  We discuss its implications to the observations of
    T-violation effects in long baseline experiments.%
}


\section{Introduction}

One of the most important questions in particle physics is whether
neutrinos are massive and if so whether there is a mixing among
leptons.  Many experiments and observations have shown evidences for
neutrino oscillation one after another, which indicates that neutrinos
are massive and there is a mixing in lepton
sector\cite{FukugitaYanagida}.  The solar neutrino deficit has long
been observed\cite{Ga1,Ga2,Kam,Cl,SolSK}.  The atmospheric neutrino
anomaly has been found\cite{ AtmKam,IMB,SOUDAN2,MACRO} and recently
almost confirmed by SuperKamiokande\cite{AtmSK}.  There is also
another suggestion given by LSND\cite{LSND}.  All of them can be
understood by neutrino oscillation.

Since there is a mixing in lepton sector, it is quite natural to
imagine that there occurs CP violation in lepton sector.  Several
physicists have considered whether we may see CP-violation effect in
lepton sector through long baseline neutrino oscillation
experiments. First it has been studied in the context of currently
planed experiments\cite{Tanimoto,ArafuneJoe,AKS,MN,BGG,GPZ,BCR,Yasuda}
and recently in the context of neutrino
factory\cite{BGW,Tanimoto2,Romanino,GH}.

In this paper we consider very long baseline experiments($\sim$ 10000
km) with low energy neutrinos ($\sim$ several hundreds
MeV)\cite{PRISM,KS2}.  We have great advantages to use them to see CP
violation effect in long baseline experiments:

\begin{enumerate}
    \item
    Availability of incident $\nu_{\rm e} (\bar\nu_{\rm e})$ beam.


    T-conjugate channel can be compared in long baseline neutrino
    oscillation experiments by use of electron neutrinos.  It enables
    to search ``matter-effect-free'' CP-violation effect.
    
    \item
    Good energy resolution and well calculated flux.

    This is also important.  We need to know neutrino flux very precisely
    since CP-violation effect may be a small effect.  Clear
    experimental signals are needed.
    
    \item
    Availability of low energy neutrinos.

    The most important point.  At least three kinds of neutrinos must
    play an important role in neutrino oscillation to search
    CP-violation effect since it arises as three(or more)-generation
    phenomena\cite{Cabibbo,BWP}.  We have to use low energy neutrinos
    to make them play the game.
    
\end{enumerate}

Thus it is of great interest to consider the possibility to observe CP
violation in long baseline experiments with low energy (several
hundreds MeV) neutrinos\cite{KS2}.

Experimentally, the observation of solar neutrino allows two regions
of mixing angle\cite{Fogli}, usually referred to as large
(MSW\cite{Wolf,MS}) and small (MSW \& Vacuum) mixing angle solution. 
It seems difficult to see CP violation for the small mixing angles,
since in such a case CP-violation effect is also in general small.  It
is worthwhile to investigate the possibility of CP violation search in
case the mixing angle is small.

From this viewpoint we consider in this paper possibilities of
observing CP-violation effects in neutrino oscillation experiments
with very long baseline ($\sim$ 10000 km) using low energy neutrino
($\sim$ several hundreds MeV).  We focus our attention to
``T-violation'' channel\cite{KuoPantaleone,KP,Toshev}.  Three
generations of neutrinos without any sterile ones are assumed, with
solar neutrino deficit and atmospheric neutrino anomaly attributed to
the neutrino oscillation.


\section{Mixing matrix and mass in the Earth}

We first derive here the mixing matrix and mass squared in the Earth
for neutrinos with low energy (several hundreds MeV). assuming three
generations of active neutrinos without any sterile one.

Let the mass eigenvalues of three neutrinos be $m_{i} (i=1, 2, 3)$ and
MNS mixing matrix\cite{MNS} $U$ which relates the flavor eigenstates
$\nu_{\alpha} (\alpha={\rm e}, \mu, \tau)$ and the mass eigenstates in
the vacuum $\nu\,'_{i} (i=1, 2, 3)$ as
\begin{equation}
  \nu_{\alpha} = U_{\alpha i} \nu\,'_{i}.
  \label{Udef}
\end{equation}
We parametrize $U$ \cite{ChauKeung,KuoPantaleone,Toshev} with
the Gell-Mann matrices $\lambda_{i}$'s as

\begin{eqnarray}
& &
U
=
{\rm e}^{{\rm i} \psi \lambda_{7}} \Gamma {\rm e}^{{\rm i} 
\phi \lambda_{5}} {\rm e}^{{\rm i} \omega \lambda_{2}} \nonumber 
\\
&=&
\left(
\begin{array}{ccc}
  1 & 0 & 0  \\
  0 & c_{\psi} & s_{\psi} \\
  0 & -s_{\psi} & c_{\psi}
\end{array}
\right)
\left(
\begin{array}{ccc}
  1 & 0 & 0  \\
  0 & 1 & 0  \\
  0 & 0 & {\rm e}^{{\rm i} \delta}
\end{array}
\right)
\left(
\begin{array}{ccc}
  c_{\phi} & 0 &  s_{\phi} \\
  0 & 1 & 0  \\
  -s_{\phi} & 0 & c_{\phi}
\end{array}
\right)
\left(
\begin{array}{ccc}
  c_{\omega} & s_{\omega} & 0 \\
  -s_{\omega} & c_{\omega} & 0  \\
  0 & 0 & 1
\end{array}
\right)
\nonumber \\
&=&
\left(
\begin{array}{ccc}
   c_{\phi} c_{\omega} &
   c_{\phi} s_{\omega} &
   s_{\phi}
  \\
   -c_{\psi} s_{\omega}
   -s_{\psi} s_{\phi} c_{\omega} {\rm e}^{{\rm i} \delta} &
   c_{\psi} c_{\omega}
   -s_{\psi} s_{\phi} s_{\omega} {\rm e}^{{\rm i} \delta} &
   s_{\psi} c_{\phi} {\rm e}^{{\rm i} \delta}
  \\
   s_{\psi} s_{\omega}
   -c_{\psi} s_{\phi} c_{\omega} {\rm e}^{{\rm i} \delta} &
   -s_{\psi} c_{\omega}
   -c_{\psi} s_{\phi} s_{\omega} {\rm e}^{{\rm i} \delta} &
   c_{\psi} c_{\phi} {\rm e}^{{\rm i} \delta}
\end{array}
\right),
\label{UPar2}
\end{eqnarray}
where $c_{\psi} = \cos \psi, s_{\phi} = \sin \phi$, etc.

The evolution equation of the flavor eigenstate of neutrinos $\nu =
(\nu_{\rm e}, \nu_{\mu}, \nu_{\tau})^{\rm T}$ that has energy $E$ in
matter with electron density $n_{\rm e}$ is given by
\begin{equation}
    - {\rm i} \frac{{\rm d} \nu}{{\rm d} x} = H \nu,
    \label{eq:evol-eq}
\end{equation}
where
\begin{eqnarray}
    H
    &=&
    U
    \left(
    \begin{array}{ccc}
        0 &   &    \\
          & \delta m^2_{21} &   \\
          &   & \delta m^2_{31}
    \end{array}
    \right)
    U^{\dagger}
    +
    \left(
    \begin{array}{ccc}
        a &   &   \\
          & 0 &   \\
          &   & 0
    \end{array}
    \right)
    \label{eq:Hdef}
    \\
    &=&
    \tilde U
    \left(
    \begin{array}{ccc}
        \tilde m^{2}_{1} &   &   \\
          & \tilde m^{2}_{2} &   \\
          &   & \tilde m^{2}_{3}
    \end{array}
    \right)
    \tilde U^{\dagger}.
    \label{eq:Hdiag}
\end{eqnarray}
Here 
\begin{equation}
    a
    =
    2 \sqrt{2} G_{\rm F} n_{\rm e} E
    =
    3.04 \times 10^{-4} {\rm eV}^{2} \cdot
    \left( \frac{E}{\rm GeV} \right)
    \left( \frac{\rho}{\rm 4.46 \ g \cdot cm^{-3}} \right)
    \label{eq:adef}
\end{equation}
is a factor due to effective mass in the matter.

The solar neutrino deficit implies $\delta m^{2} \sim O(10^{-4} \sim
10^{-5}) {\rm eV^{2}}$, while the atmospheric neutrino anomaly
suggests $\delta m^{2} \sim O(10^{-2} \sim 10^{-3}) {\rm eV^{2}}$.  We
regard them as $\delta m^{2}_{21}$ and $\delta m^{2}_{31}$
respectively.  With these values and eq.(\ref{eq:adef}) we assume $a,
\delta m^{2}_{21} \ll \delta m^{2}_{31}$ and diagonalize eq.
(\ref{eq:Hdef}) up to the lowest order.
Letting
\begin{eqnarray}
    H'
    &\equiv&
    {\rm e}^{{\rm i} \omega \lambda_{2}} U^{\dagger}
    H
    U {\rm e}^{- {\rm i} \omega \lambda_{2}}
    \nonumber \\
    &=&
    \left(
    \begin{array}{ccc}
        0 &   &   \\
          & 0 &   \\
          &   & \delta m^{2}_{31}
    \end{array}
    \right)
    \nonumber \\
    &+&
    {\rm e}^{{\rm i} \omega \lambda_{2}}
    \left(
    \begin{array}{ccc}
        0 &  &   \\
         & \delta m^{2}_{21} &    \\
         &  & 0
    \end{array}
    \right)
    {\rm e}^{- {\rm i} \omega \lambda_{2}}
    +
    {\rm e}^{{\rm i} \omega \lambda_{2}} U^{\dagger}
    \left(
    \begin{array}{ccc}
        a &   &    \\
          & 0 &    \\
          &   & 0
    \end{array}
    \right)
    U {\rm e}^{- {\rm i} \omega \lambda_{2}}
    \nonumber \\
    &=&
    H'_{0} + H'_{1},
\end{eqnarray}
where
\begin{equation}
    H'_{0} = 
    \left(
    \begin{array}{ccc}
        0 &   &   \\
          & 0 &   \\
          &   & \delta m^{2}_{31}
    \end{array}
    \right)
    \label{eq:H'0def}
\end{equation}
and
\begin{eqnarray}
    H'_{1}
    &=&
    {\rm e}^{{\rm i} \omega \lambda_{2}}
    \left(
    \begin{array}{ccc}
        0 &  &   \\
         & \delta m^{2}_{21} &    \\
         &  & 0
    \end{array}
    \right)
    {\rm e}^{- {\rm i} \omega \lambda_{2}}
    +
    {\rm e}^{{\rm i} \omega \lambda_{2}} U^{\dagger}
    \left(
    \begin{array}{ccc}
        a &   &    \\
          & 0 &    \\
          &   & 0
    \end{array}
    \right)
    U {\rm e}^{- {\rm i} \omega \lambda_{2}}
    \nonumber \\
    &=&
    \delta m^{2}_{21}
    \left(
    \begin{array}{ccc}
        s_{\omega}^{2} & c_{\omega} s_{\omega} & 0 \\
        c_{\omega} s_{\omega} & c_{\omega}^{2} & 0 \\
        0 & 0 & 0
    \end{array}
    \right)
    +
    a
    \left(
    \begin{array}{ccc}
        c_{\phi}^{2} & 0 & c_{\phi} s_{\phi} \\
        0 & 0 & 0 \\
        c_{\phi} s_{\phi} & 0 & c_{\phi}^{2}
    \end{array}
    \right),
\end{eqnarray}
we diagonalize $H'$ by perturbation.  $H'_{0}$ has degenerate 
eigenvalues, so we first diagonalize
\begin{eqnarray}
    h'_{1}
    &\equiv&
    \left(
    \begin{array}{cc}
        a c_{\phi}^{2} + \delta m^{2}_{21} s_{\omega}^{2} &
        \delta m^{2}_{21} c_{\omega} s_{\omega}
        \\
        \delta m^{2}_{21} c_{\omega} s_{\omega} &
        \delta m^{2}_{21} c_{\omega}^{2}
    \end{array}
    \right)
    \nonumber \\
    &=&
    \frac{a c_{\phi}^{2} + \delta m^{2}_{21}}{2}
    +
    \frac{1}{2}
    \left(
    \begin{array}{cc}
        a c_{\phi}^{2} - \delta m^{2}_{21} c_{2 \omega} &
        \delta m^{2}_{21} s_{2 \omega}
        \\
        \delta m^{2}_{21} s_{2 \omega} &
        -a c_{\phi}^{2} + \delta m^{2}_{21} c_{2 \omega}
    \end{array}
    \right).
    \label{eq:h1'def}
\end{eqnarray}
The eigenvalues of (\ref{eq:h1'def}) are
\begin{equation}
    \lambda_{\pm}
    =
    \frac{a c_{\phi}^{2} + \delta m^{2}_{21}}{2}
    \pm
    \frac{1}{2}
    \sqrt{
    (a c_{\phi}^{2} - \delta m^{2}_{21} c_{2 \omega})^{2} +
    (\delta m_{21}^{2})^{2} s_{2 \omega}^{2}
    },
    \label{eq:lambda+-def}
\end{equation}
so that $h'_{1}$ is diagonalized as
\begin{equation}
    h'_{1}
    =
    \left(
    \begin{array}{cc}
        \cos \tilde \omega & \sin \tilde \omega \\
        -\sin \tilde \omega & \cos \tilde \omega
    \end{array}
    \right)
    \left(
    \begin{array}{cc}
        \lambda_{-} & 0 \\
        0 & \lambda_{+} \\
    \end{array}
    \right)
    \left(
    \begin{array}{cc}
        \cos \tilde \omega & - \sin \tilde \omega \\
        \sin \tilde \omega & \cos \tilde \omega
    \end{array}
    \right)
    \label{eq:h1'diag}
\end{equation}
where $\tilde \omega$ satisfies
\begin{eqnarray}
    \tan 2 \tilde \omega
    &=&
    \frac{\delta m^{2}_{21} s_{2 \omega}}
    {- a c_{\phi}^{2} + \delta m^{2}_{21} c_{2 \omega}},
    \label{eq:tan2omegatilde} \\
    \sin 2 \tilde \omega
    &=&
    \frac{\delta m^{2}_{21} s_{2 \omega}}
    {
        \sqrt{
            (a c_{\phi}^{2} - \delta m^{2}_{21} c_{2 \omega})^{2} +
            (\delta m^{2}_{21} s_{2 \omega})^{2}
        }
    }.
    \label{eq:sin-tilde-omega}
\end{eqnarray}
%

%
The eigenvalues of $H'$ (and thus of $H$) to the first order are thus
\begin{eqnarray}
    \tilde m^{2}_{1}
    &=&
    \frac{a c_{\phi}^{2} + \delta m^{2}_{21}}{2}
    -
    \frac{1}{2}
    \sqrt{
    (a c_{\phi}^{2} - \delta m^{2}_{21} c_{2 \omega})^{2} +
    (\delta m_{21}^{2})^{2} s_{2 \omega}^{2}
    },
    \label{eq:m2tilde1} \\
    \tilde m^{2}_{2}
    &=&
    \frac{a c_{\phi}^{2} + \delta m^{2}_{21}}{2}
    +
    \frac{1}{2}
    \sqrt{
    (a c_{\phi}^{2} - \delta m^{2}_{21} c_{2 \omega})^{2} +
    (\delta m_{21}^{2})^{2} s_{2 \omega}^{2}
    },
    \label{eq:m2tilde2}
\end{eqnarray}
and
\begin{equation}
    \tilde m^{2}_{3}
    =
    \delta m^{2}_{31} + a s_{\phi}^{2}.
    \label{eq:m2tilde3}
\end{equation}
%

We finally obtain at the lowest order
\begin{equation}
    H =
    [ U {\rm e}^{- {\rm i}
    (\omega - \tilde \omega) \lambda_{2}}
    ]
    \cdot
    {\rm diag} (\tilde m^{2}_{1}, \tilde m^{2}_{2}, \tilde m^{2}_{3})
    \cdot
    [ U {\rm e}^{- {\rm i}
    (\omega - \tilde \omega) \lambda_{2}}
    ]^{\dagger}.
    \label{eq:Hdiag2}
\end{equation}

Note that (\ref{eq:sin-tilde-omega}) has resonant peak at $\tilde
\omega = \pi / 4$ or
\begin{equation}
    a c_{\phi}^{2} = \delta m^{2}_{21} c_{2 \omega}.
    \label{eq:omega-resonance}
\end{equation}
This condition can be written in terms of neutrino energy $E$ as
\begin{equation}
    E
    =
    E_{\rm peak}
    \equiv
    294 {\rm MeV}
    \times
    \frac{c_{2 \omega}}{c_{\phi}^{2}}
    \left( \frac{\delta m^{2}_{21}}{10^{-4}{\rm eV^{2}}} \right)
    \left( \frac{\rho}{\rm 4.46 \ g \cdot cm^{-3}} \right)^{-1}.
    \label{eq:E-peak}
\end{equation}
It is quite interesting that $\sin 2 \tilde \omega$ grows up to
maximum value of 1 when $E$ has certain value determined by
eq.(\ref{eq:omega-resonance}) or eq.(\ref{eq:E-peak}), regardless how
small $\omega$ itself may be.
We persue in the next section the possible merit of this enhancement in the 
CP or T violation search with the long baseline neutrino oscillation 
experiments.


\section{T-violation search in the very long baseline experiments}

\subsection{Resonance of T-violation effect}

The factor
\begin{equation}
    U {\rm e}^{{\rm -i} (\omega - \tilde \omega)}
    =
    {\rm e}^{{\rm i} \psi \lambda_{7}}
    \Gamma
    {\rm e}^{{\rm i} \phi \lambda_{5}}
    {\rm e}^{{\rm i} \tilde \omega \lambda_{2}},
    \label{eq:Utilde}
\end{equation}
which appears in eq.(\ref{eq:Hdiag2}), is different from $U$ in
eq.(\ref{UPar2}) only by a simple replacement $\omega \rightarrow
\tilde \omega$.  So the matter effect is included in the lowest order
by the substitution $\delta m^{2}_{ij} \rightarrow \delta \tilde
m^{2}_{ij} \equiv \tilde m^{2}_{i} - \tilde m^{2}_{j}$ and $\omega
\rightarrow \tilde \omega$.  The T-violation effect in the presence of
matter is thus given in this order by
\begin{equation}
    P(\nu_{\mu} \rightarrow \nu_{\rm e}) -
    P(\nu_{\rm e} \rightarrow \nu_{\mu})
    =
    4 \tilde J
    \left(
    \sin \frac{\delta \tilde m^{2}_{21} L}{2 E} +
    \sin \frac{\delta \tilde m^{2}_{32} L}{2 E} +
    \sin \frac{\delta \tilde m^{2}_{13} L}{2 E}
    \right),
    \label{eq:Tviol-in-matter}
\end{equation}
where
\begin{eqnarray}
    \tilde J
    &=&
    \cos^{2} \phi \sin \phi
    \cos \psi \sin \psi
    \cos \tilde \omega \sin \tilde \omega
    \sin \delta
    \nonumber \\
    &=&
    \frac{1}{4}
    \cos^{2} \phi \sin \phi \sin 2 \psi \sin 2 \tilde \omega \sin \delta
    \label{eq:Jtilde}
\end{eqnarray}
is the effective Jarlskog parameter in matter.  We have already stated
at the end of previous section that $\sin 2 \tilde \omega$ has
resonant peak at certain energy.  $\tilde J$ has indeed this factor,
so the T-violation effect is expected to have resonant peak.

\begin{figure}
 \unitlength=1cm
 \begin{picture}(15,18)
  \unitlength=1mm
  \centerline{
   \epsfysize=18cm
   \epsfbox{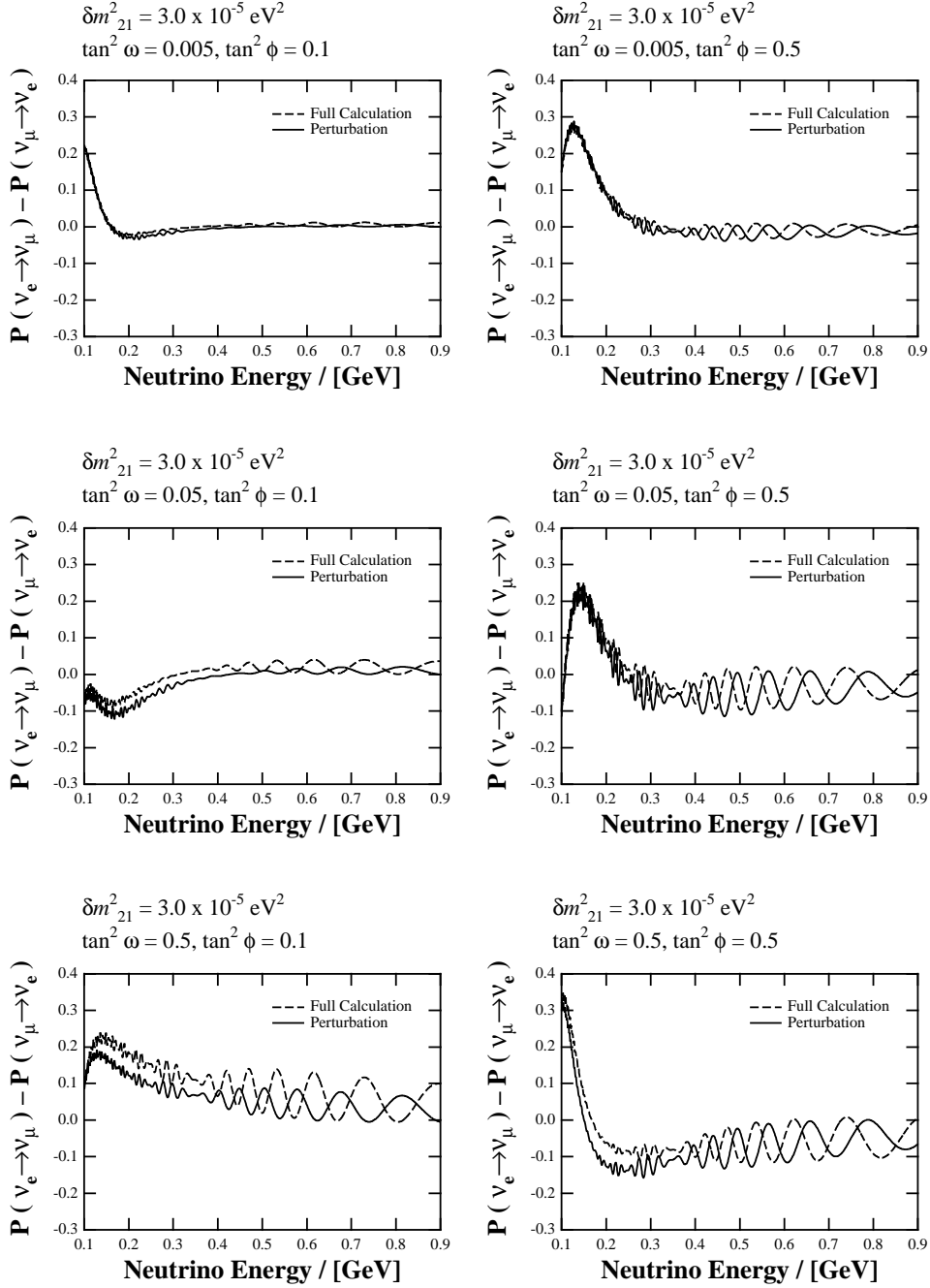}
  }
 \end{picture}
\caption{%
	The approximated oscillation probabilities (solid lines)
	compared with unapproximated ones (dashed lines).  The
	parameters not indicated in the figures are taken same as
	eqs.(\ref{eq:example-params}).  Some parameter sets shown here are 
	not favored by the solar neutrino observations, but we have 
	presented them together to show the tendency of the graphs.%
	}
\label{fig:1e-3-3e-5}
\end{figure}

\begin{figure}
 \unitlength=1cm
 \begin{picture}(15,18)
  \unitlength=1mm
  \centerline{
   \epsfysize=18cm
   \epsfbox{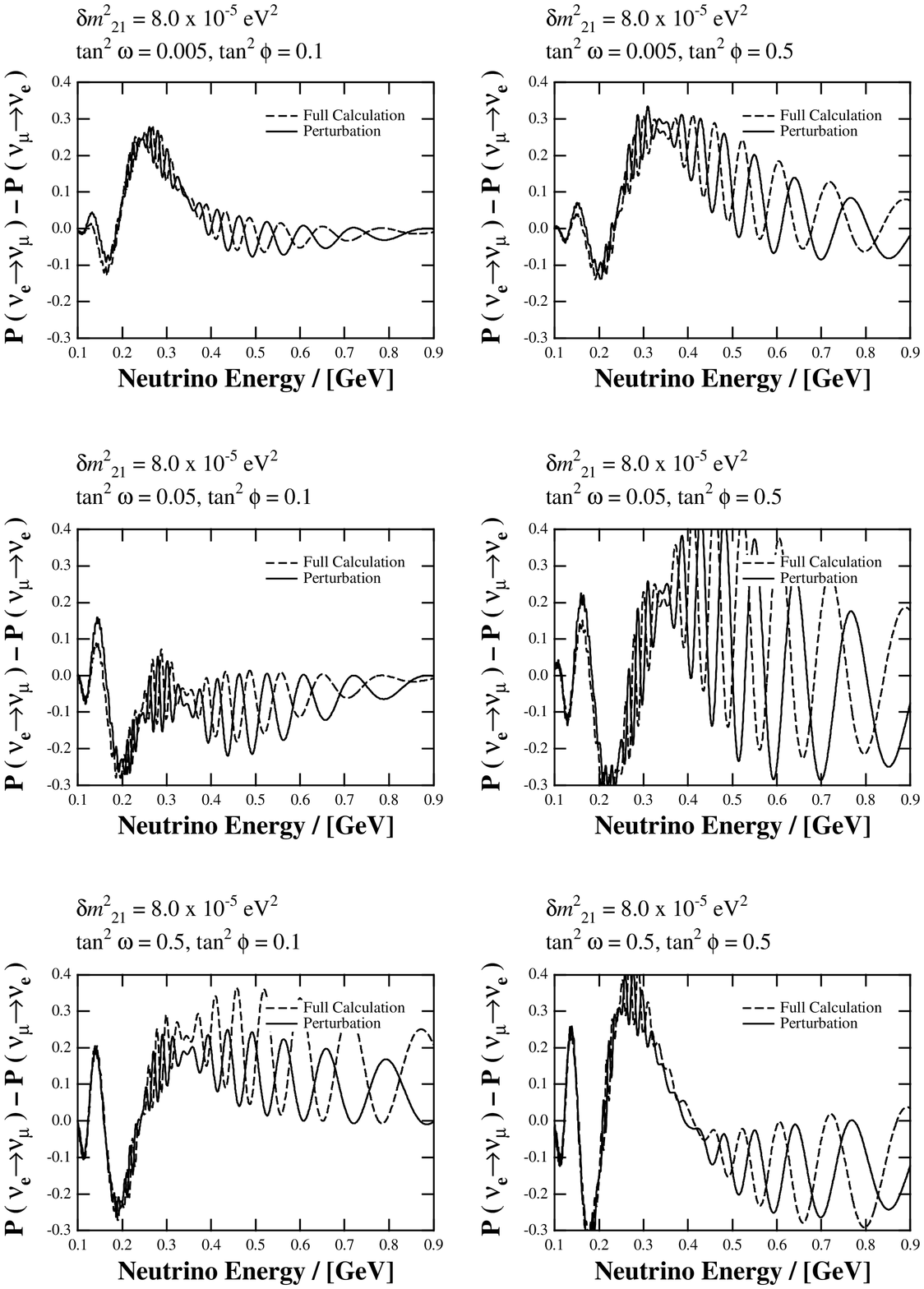}
  }
 \end{picture}
\caption{%
        Same as Fig.\ref{fig:1e-3-8e-5}, but $\delta m^{2}_{21} = 8 \times 
        10^{-5} {\rm eV}^{2}$.  }
\label{fig:1e-3-8e-5}
\end{figure}

With this prospect we present in Fig.\ref{fig:1e-3-3e-5} and
Fig.\ref{fig:1e-3-8e-5} the T-violation effect averaged over energy
resolution (here tentatively taken as 30 MeV) as a function of energy
$E$.  Both the perturbed result and the full calculation result are
shown to exhibit how well our approximation works.  The parameters not
shown in the figure are fixed as an example to
\begin{equation}
    \delta m^{2}_{31} = 1 \times 10^{-3} {\rm eV^{2}},\;
    \psi = \frac{\pi}{4},\;
    L = 10000 {\rm km},
    \quad {\rm and} \quad
    \rho = 4.46 {\rm g / cm^{-3}}.
    \label{eq:example-params}
\end{equation}
$L$ is taken to 10000 km since it must be the same order to the
longest oscillation length,
\begin{equation}
    l_{21} \equiv
    \frac{4 \pi E}{\delta m^{2}_{21}} =
    2.48 \times 10^{4} {\rm km}
    \cdot
    \left( \frac{E}{\rm GeV} \right)
    \cdot
    \left( \frac{\delta m^{2}_{21}}{\rm 10^{-4} {\rm eV^{2}}} \right)^{-1}.
    \label{eq:osclen-21}
\end{equation}
The value of $\rho$ in eq.(\ref{eq:example-params}) is the average
value of the crust and mantle, allowing that $L = 10000$ km can be
realized without touching core of the Earth.


\begin{figure}
 \unitlength=1cm
 \begin{picture}(15,18)
  \unitlength=1mm
  \centerline{
   \epsfysize=18cm
   \epsfbox{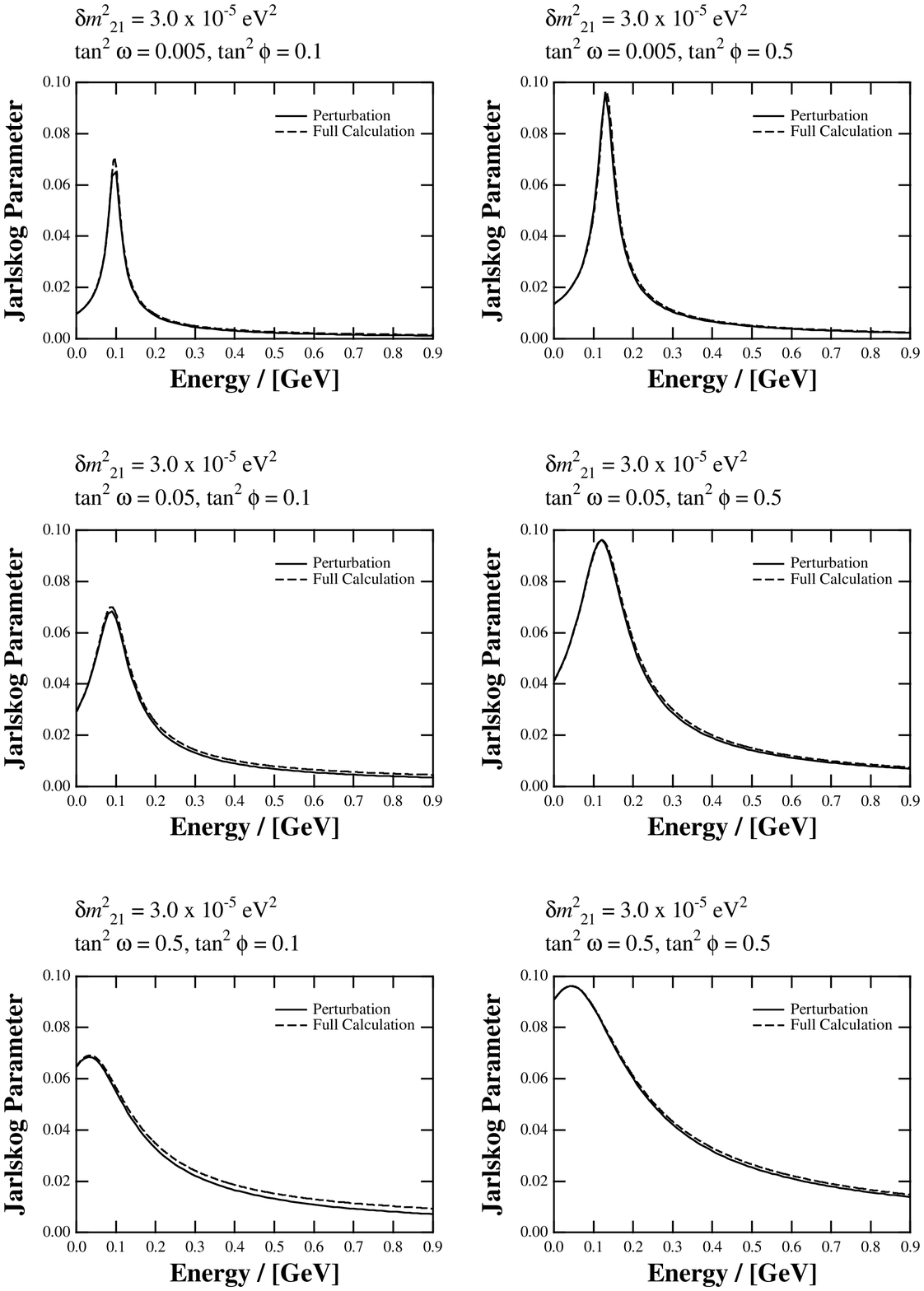}
  }
 \end{picture}
\caption{%
        Effective Jarlskog parameter in the matter as a function of neutrino 
        energy.  Parameters are taken same as in Fig.\ref{fig:1e-3-3e-5}.}
\label{fig:Jtilde-3e-5}
\end{figure}

\begin{figure}
 \unitlength=1cm
 \begin{picture}(15,18)
  \unitlength=1mm
  \centerline{
   \epsfysize=18cm
   \epsfbox{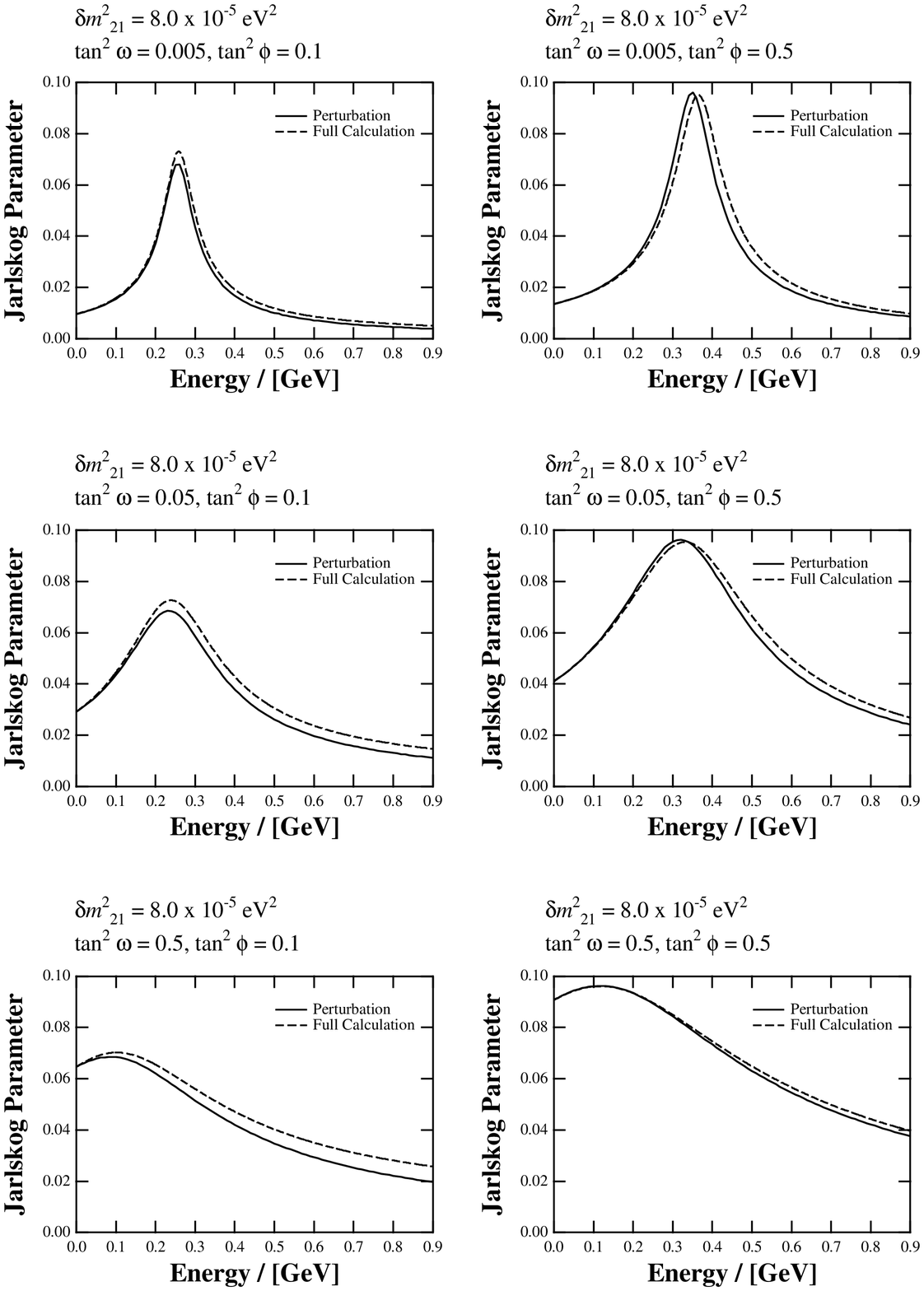}
  }
 \end{picture}
\caption{%
        Same as Fig.\ref{fig:Jtilde-3e-5}, but parameters are same as in 
        Fig.\ref{fig:1e-3-3e-5}. }
\label{fig:Jtilde-8e-5}
\end{figure}

The graphs of $\tilde J$ as a function of energy, both
perturbed and full calculation results, is also displayed in
Fig.  \ref{fig:Jtilde-3e-5} and Fig.  \ref{fig:Jtilde-8e-5}.

We observe in the figures that our approximation works well, and as we
stated both the T-violation effect and $\tilde J$ has resonant peak. 
The T-violation effect gets up over 10\% for some case even $\omega$
is considerably small thanks to the resonant peak of $\tilde J$,
leaving chance to search T-violation effect for such small-$\omega$
region.

Note that not only $\tilde J$ but also summation of three $\sin (\delta \tilde
m^{2}_{ij} L) / (2 E)$'s must be large in order to obtain large
T-violation effect.  Thus the peak of T violation may shift a little
from $E_{\rm peak}$.


\subsection{The limits of the T-violation search in terms of $\sin^{2}
2 \omega$ and $\delta m^{2}_{21}$}

We consider here limits of parameter region we can cover in search of
T violation.  We will give limits in terms of $\delta m^{2}_{21}$ and 
$s_{2 \omega}$ so that we can refer to the diagram of the solar neutrino 
drawn in $\sin^{2} 2 \theta$-$\delta m^{2}$ plane.

We first pay attention to the behavior of $\tilde J$ near $E = E_{\rm
peak}$ referring eq.(\ref{eq:sin-tilde-omega}).  The half-width
$\Delta a$ of the resonance of $\sin \tilde \omega$ can be estimated
from
\begin{equation}
    \Delta a \cdot c_{\phi}^{2} = \delta m^{2}_{21} s_{2 \omega}.
    \label{eq:a-hw}
\end{equation}
The half-width in terms of the neutrino energy, $\Delta E$, is thus
given by
\begin{eqnarray}
    \Delta E
    & = &
    \frac{\delta m^{2}_{21}}{2 \sqrt{2} G_{\rm F} n_{\rm e}}
    \frac{s_{2 \omega}}{c_{\phi}^{2}}
    \nonumber \\
    & = &
    294 {\rm MeV}
    \times
    \frac{s_{2 \omega}}{c_{\phi}^{2}}
    \left( \frac{\delta m^{2}_{21}}{10^{-4}{\rm eV^{2}}} \right)
    \left( \frac{\rho}{\rm 4.46 g\ cm^{-3}} \right)^{-1}.
    \label{eq:resonance-width}
\end{eqnarray}
The resonance gets sharper and narrower as $\sin 2 \omega$ gets
smaller.  We can pick up the resonance peak if the width is large
compared the energy resolution $\delta E$.  The condition $\Delta E \ge
\delta E$ leads to
\begin{equation}
   \delta m^{2}_{21} s_{2 \omega}
   \ge
   10^{-4} {\rm eV^{2}} \cdot
   c_\phi^2 \left( \frac{\rho}{\rm 4.46 g\ cm^{-3}} \right)
   \left( \frac{\delta E}{294 {\rm MeV}} \right).
\end{equation}
As the width gets smaller than the
resolution the observed T violation effect will be averaged over
outside the peak, so we can make less use of the resonance.  Thus the 
smaller energy resolution enables to reach smaller $\omega$ region in 
search of T-violation effect.

There is another limit in searching small $\omega$ region.  
We need
\begin{equation}
    \frac{\delta \tilde m^{2}_{21} L}{2 E}
    \stackrel{>}{\sim}
    1
    \quad {\rm at} \quad
    E = E_{\rm peak},
    \label{eq:phase-cond}
\end{equation}
so that we make use of the resonance peak.  Equation 
(\ref{eq:phase-cond}) leads to 
\begin{equation}
    \frac{2 \pi L}{l_{21}} s_{2 \omega}
    \stackrel{>}{\sim}
    1
    \label{eq:phase-cond2}
\end{equation}
or, with eqs. (\ref{eq:m2tilde1}), (\ref{eq:m2tilde2}) and 
(\ref{eq:osclen-21}),
\begin{equation}
    \delta m^{2}_{21} s_{2 \omega}
    \stackrel{>}{\sim}
    0.39 \times 10^{-4} {\rm eV^{2}}
    \cdot
    \left( \frac{E}{\rm GeV} \right)
    \left( \frac{L}{10000 {\rm km}} \right)^{-1}.
    \label{eq:phase-cond3}
\end{equation}
%
Almost all part of large MSW mixing angle solution can be thus covered
with this result.  Note that one can reach still smaller value by
taking the tail of the peak at $E = E_{\rm peak}$.  For instance, the
limit in case one observe T-violation effect at $E = E_{\rm peak} \pm
\Delta E$ is
\begin{equation}
    \delta m^{2}_{21} s_{2 \omega}
    \stackrel{>}{\sim}
    0.28 \times 10^{-4} {\rm eV^{2}}
    \cdot
    \left( \frac{E}{\rm GeV} \right)
    \left( \frac{L}{10000 {\rm km}} \right)^{-1}.
    \label{eq:phase-cond'}
\end{equation}

\begin{figure}
 \unitlength=1cm
 \begin{picture}(15,18)
  \unitlength=1mm
  \centerline{
   \epsfysize=18cm
   \epsfbox{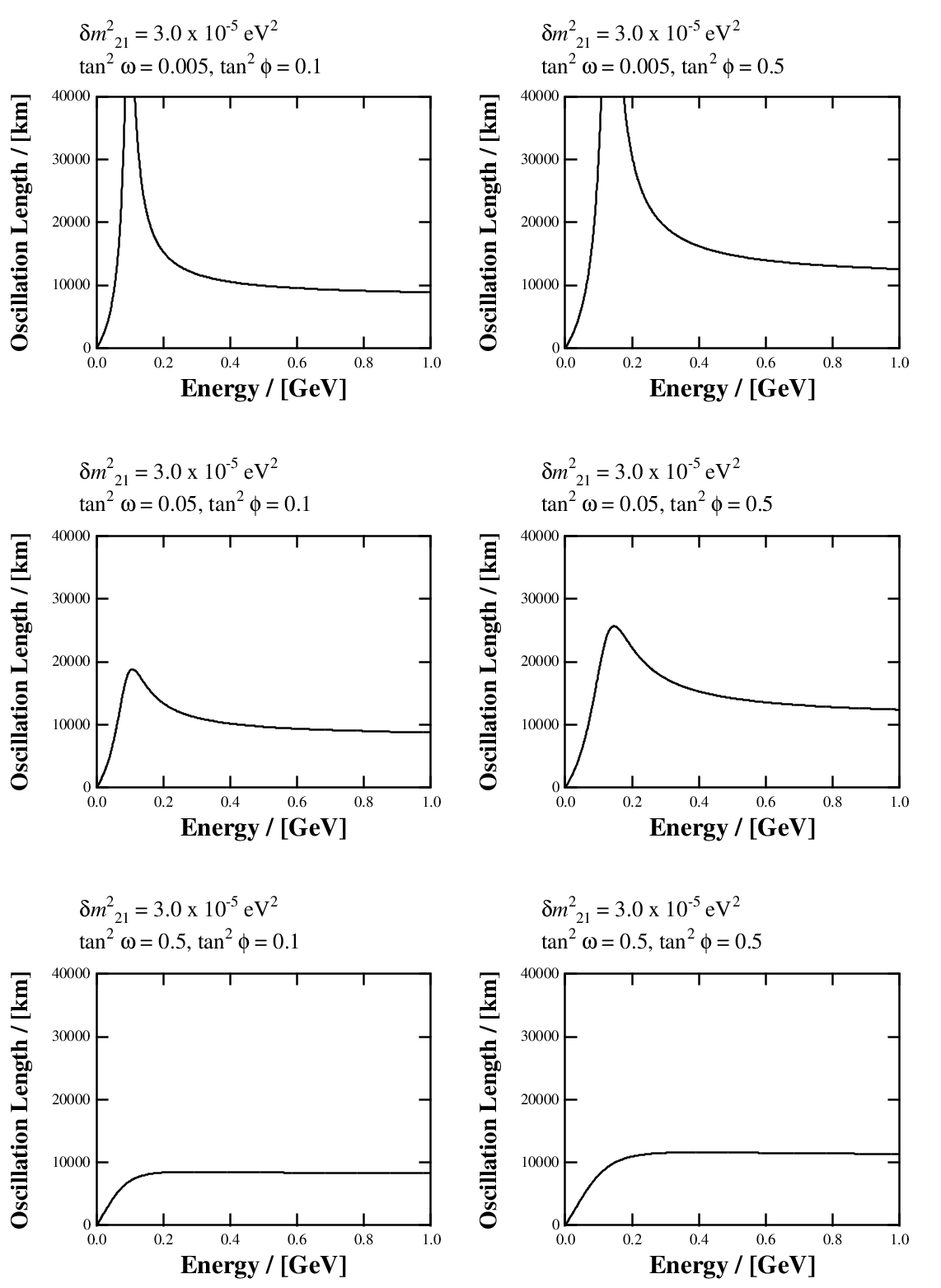}
  }
 \end{picture}
\caption{%
        The longest oscillation length $\tilde l_{21}$ in the matter
        as a function of neutrino energy.  The graph is calculated 
        from the perturbated results.  Parameters are taken same
        as in Fig.\ref{fig:osclen-3e-5}.}
\label{fig:osclen-3e-5}
\end{figure}

\begin{figure}
 \unitlength=1cm
 \begin{picture}(15,18)
  \unitlength=1mm
  \centerline{
   \epsfysize=18cm
   \epsfbox{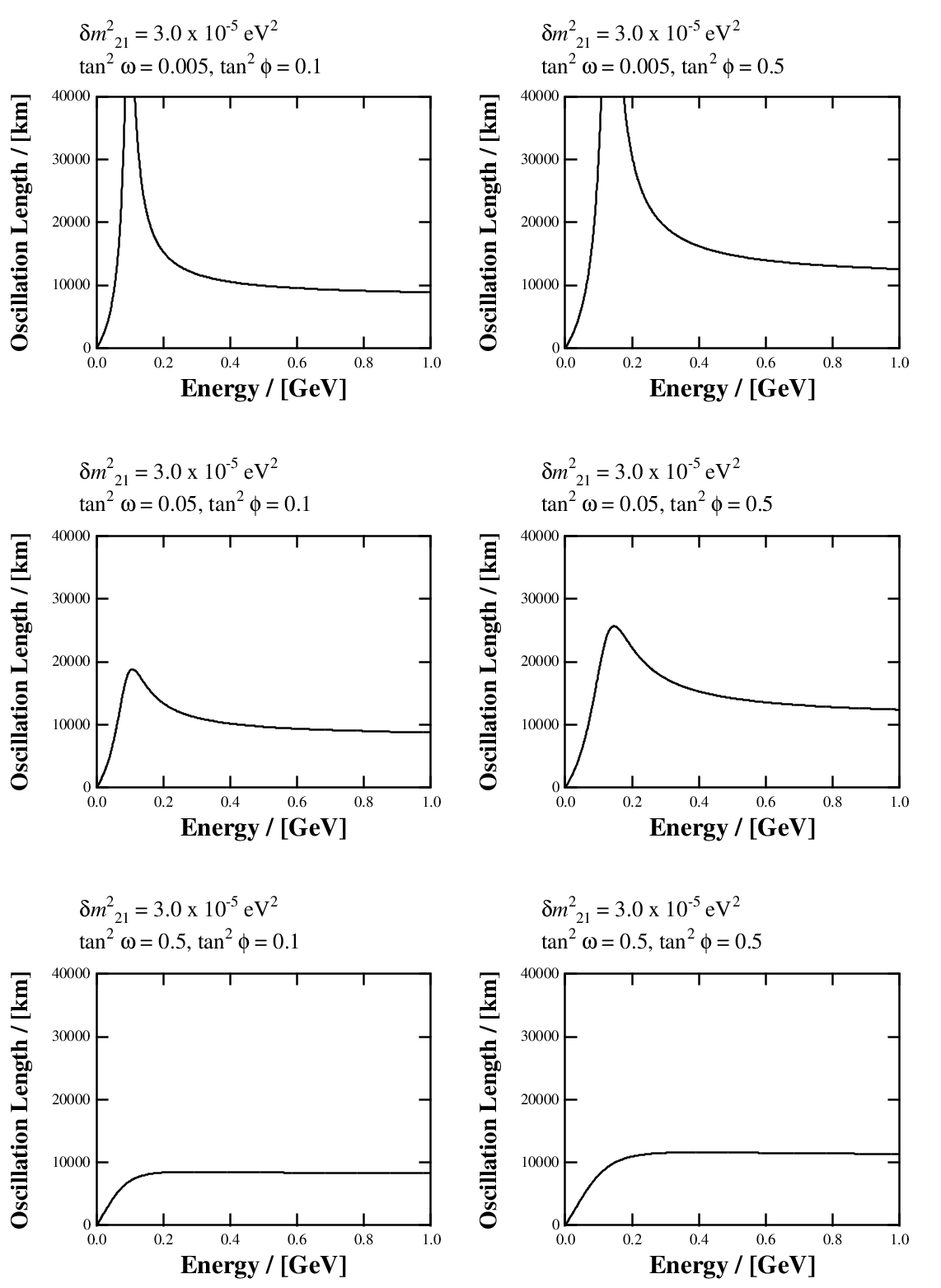}
  }
 \end{picture}
\caption{%
        Same as Fig.\ref{fig:osclen-3e-5}, but parameters are same as in 
        Fig.\ref{fig:1e-3-3e-5}. }
\label{fig:osclen-8e-5}
\end{figure}

In Fig.\ref{fig:osclen-3e-5} and Fig.\ref{fig:osclen-8e-5} we show
how the oscillation length $\tilde l_{21}$ changes as a function of
neutrino energy calculated from the perturbation result.


\section{Summary and discussion}
We have considered the possibility of searching T violation by the
very long baseline neutrino oscillation experiments.  The matter on
the baseline effectivly changes only the mixing angle of the
first-second generation.  The effective Jarlskog parameter has an
resonant peak at $\sim O(100 {\rm MeV})$ and also T violation effect
grows by resonance.  Making use of the resonance, the wider region in
$\delta m^{2}_{21}$-$s^{2}_{2 \omega}$ plane can be covered in the
search of T violation.  The limits of search is given by the energy
resolution and baseline length.  Almost all part of the large MSW
mixing angle solution can be searched with $L \sim 10000$ km and $E
\sim O(100 {\rm MeV})$.


\end{document}